\documentclass[twocolumn]{aastex6}
\usepackage{aas_macros}
\usepackage{todonotes}
\usepackage{graphicx}
\usepackage{enumerate}
\usepackage{color}
\usepackage{amsmath}
\usepackage{amssymb}
\usepackage{hyperref}
\usepackage{xspace}
\usepackage{booktabs}
\usepackage{subcaption}

\newcommand{\skyfact}{{\textsc{SkyFACT}}}
\newcommand{\run}{{\textsc{Run}}}
\newcommand{\Fermi}{\textit{Fermi}}
\newcommand{\gr}{$\gamma$-ray\xspace}

\received{\today}
\revised{}
\accepted{}

\shorttitle{The Fermi-LAT GCE Traces Stellar Mass in the Bulge}
\shortauthors{R. Bartels et al.}

\begin{document}

\title{The Fermi-LAT GeV Excess Traces Stellar Mass in the Galactic Bulge}

\email{r.t.bartels@uva.nl, c.weniger@uva.nl, e.storm@uva.nl, calore@lapth.cnrs.fr}

\author{Richard Bartels, Emma Storm \& Christoph Weniger}
\affiliation{GRAPPA, University of Amsterdam\\
Science Park 904,
1090GL Amsterdam, The Netherlands}

\author{Francesca Calore}
\affiliation{LAPTh, CNRS \\
9 Chemin de Bellevue, 74941 Annecy-le-Vieux, France}

\begin{abstract}
An anomalous emission component at energies of a few GeV and located towards
the inner Galaxy is present in the \Fermi-LAT data. It is known as the
\Fermi-LAT GeV excess.  Using almost 8 years of data we reanalyze the
characteristics of this excess with \skyfact, a novel tool that combines image
reconstruction with template fitting techniques.  We find that an emission
profile that traces stellar mass in the boxy and nuclear bulge provides the
best description of the excess emission, providing strong circumstantial
evidence that the excess is due to a stellar source population in the Galactic
bulge.  We find a luminosity to stellar mass ratio of $(2.1\pm 0.2)\times
10^{27} \mathrm{\,erg\,s^{-1}\,M_\odot^{-1}}$ for the boxy bulge, and of
$(1.4\pm 0.6)\times 10^{27}\mathrm{\,erg\,s^{-1}\,M_\odot^{-1}}$ for the
nuclear bulge.  Stellar mass related templates are preferred over conventional
DM profiles with high statistical significance.  
\end{abstract}



\section{Introduction}
\label{sec:intro}

An anomalous emission component, often referred to as the Galactic center GeV
excess (GCE), has been identified in the \Fermi-LAT data by many groups
\citep[e.g.~][]{Goodenough:2009gk, Vitale:2009hr, Hooper:2011ti,
Abazajian:2012pn, Macias:2013vya, Daylan:2014rsa, Zhou:2014lva, Calore:2014xka,
Huang:2015rlu, deBoer:2016esu, TheFermi-LAT:2015kwa}.  Its spectrum peaks at
energies of a few GeV and it appears to be uniform over the emission region.
The morphology is usually described as almost spherically symmetric around the
Galactic center, with a radial extent of $\sim 10^\circ$.  Intriguingly, a
signal from dark matter (DM) annihilation into $b$-quark pairs and a DM mass
$\sim 50\rm\,GeV$ has been shown to be consistent with the GCE
\citep{Goodenough:2009gk, Abazajian:2012pn, Macias:2013vya, Daylan:2014rsa,
Calore:2014nla}, provided the centrally peaked DM distribution in the Galactic
bulge follows a radial power-law profile with index $\gamma \sim 1.2$.
However, the exact details of the morphology and spectrum remain subject to
debate, in particular due to the uncertainties in the interstellar emission
modeling \citep{Carlson:2016iis, TheFermi-LAT:2017vmf}.  Additionally, there is
strong degeneracy with the \Fermi\ Bubbles, giant diffuse lobes oriented
perpendicularly to the Galactic plane
\citep{2010ApJ...717..825D,2010ApJ...724.1044S,Fermi-LAT:2014sfa}, the
low-latitude behavior of which is not well-characterized
\citep{TheFermi-LAT:2017vmf,Linden:2016rcf}.

Besides DM, more 'conventional' astrophysical
explanations do exist, with various degrees of plausibility.  These are either
related to a large number of hitherto unresolved point sources in the Galactic
bulge, just at and below the detection threshold of \Fermi-LAT, or to
diffuse photons coming from a central population of cosmic rays.  Nowadays, a
population of unresolved millisecond pulsars (MSPs), whose \gr spectrum was
shown to match that of the GCE \citep{Abazajian:2010zy, Abazajian:2014fta,
Calore:2014xka}, represents the most promising astrophysical
interpretation to the GCE~\citep{Abazajian:2010zy, Gordon:2013vta,
Petrovic:2014xra, Yuan:2014rca}.  Corroborative evidence for this
interpretation was recently found in analyses of the \gr\ data using wavelet
fluctuations, and non-Poissonian template fits \citep{Bartels:2015aea,
Lee:2015fea}.  
Spectral classification of low-significance \gr\ sources and 
analyses of their distribution 
remain however inconclusive about the presence of a bulge population 
\citep{Fermi-LAT:2017yoi,Bartels:2017xba}.
Lastly, it was recently shown
that deep learning is another potentially powerful tool to discriminate between a point-source-like or diffuse-like structure of the excess \citep{Caron:2017udl}. Although the evidence for the point source
scenario is growing, it will require follow-up radio observations with
MeerKAT and/or SKA to robustly confirm this interpretation
\citep{Calore:2015bsx}, and exclude a DM interpretation for good. 

Nevertheless, arguments against MSPs in the bulge exist and are based, for example, on 
the discrepancy between the required MSPs and the observed number of low-mass X-ray binaries
progenitors~\citep{Haggard:2017lyq,Cholis:2014fja} or on an implausibly high formation efficiency of MSPs in globular clusters~\citep{Hooper:2016rap}. 
However, MSP evolutionary channels are complex and remain highly uncertain~\citep{Ploeg:2017vai}. 

Surprisingly, possible connections between the morphology of the GCE and the
morphology of the observed Galactic bulge received only little attention in the
literature.  The Milky Way hosts a central boxy/peanut-shaped bulge/bar that
was likely formed from the buckling instability after the bar formation through
bar instability \citep[see e.g.~][]{2016ASSL..418..233S}.  The stellar mass of
this boxy/peanut bulge, which is mostly made of old ($>$5 Gyr) stellar
populations, is estimated to be $\sim 10^{10}\mathrm{\,M_\odot}$,
\citep[e.g.~][]{Cao:2013dwa, Portail:2016vei} about 15\% of the total stellar
mass in the Galaxy \citep{McMillan:2011wd,Licquia:2014rsa}.  The Galactic bulge
has a radial extension of about 3 kpc and shows a complex morphological
structure both in its stellar and gas content.  It transitions into a thinner
long bar component which extends about $\sim5\mathrm{\,kpc}$
\citep{2015MNRAS.450.4050W}.  In the innermost $\sim$ 200 pc, we find the
nuclear bulge (NB), a region of very high stellar density consisting of the
nuclear stellar disk and the nuclear stellar (or star) cluster
\citep{Launhardt:2002tx,Portail:2016vei}.  In addition to the boxy/pseudo
bulge, there exists evidence for the presence of a spherical classical bulge,
revealed through metal-poor RR-Lyrae stars
\citep{Dekany:2013ipa,2016ApJ...821L..25K}.  This component is only expected to
contribute $1\%$ to the total mass in the inner-Galaxy, compared to $\sim90\%$
for the boxy bulge \citep{2016ApJ...821L..25K}.  Finally, there is evidence for
an X-shaped component \citep{McWilliam:2010bc, Nataf:2010wf}, which can
naturally form from the buckling instability and, as such, is a product of bar
evolution \citep{Li:2012qr}.  Estimates of the mass of the X-shaped bulge range
from a few percent of the total bulge mass \citep{Li:2012qr, Cao:2013dwa} up to
$\sim 45\%$ \citep{2015MNRAS.450L..66P}.  Recently, \cite{Macias:2016nev}
claimed that the GCE traces this X-shaped bulge and in the very
center the nuclear bulge.

\medskip

In the present paper, we analyze the GCE using our newly developed code
\skyfact\ (Sky Factorization with Adaptive Constrained Templates),
\citealt{Storm:2017arh}, which is a hybrid approach between template fitting and image
reconstruction and allows for a much larger range of modeling
systematics than previously possible.  We compare the morphology of the GCE not
only to DM-inspired models, but also to models that fit the stellar
distribution in the inner Galaxy.  The paper is organized as follows:  in
Section~\ref{sec:modeling} we describe our analysis methodology, using
\skyfact.  Results are presented and discussed in Section~\ref{sec:results},
while we present our conclusions in Section~\ref{sec:conclusions}. Full methods
and a summary of \skyfact\ are presented in Appendix~\ref{sec:methods}. In the
supplemental material, Appendix~\ref{sec:supp}, we describe a number of
systematics that could affect our results, including  a detailed comparison of
\skyfact\ results and previous analyses, and the effects of the \Fermi\
bubbles, additional point sources and star formation in the inner Galaxy.

\section{Modeling the \gr sky}
\label{sec:modeling}

We model the \gr\ sky using \skyfact, a newly developed code for \gr\ data
analysis that, through a combination of image reconstruction techniques and template fitting,
accounts for expected spatial and spectral uncertainties in the various emission
components by allowing a large number of `nuisance parameters'
\citep{Storm:2017arh}.  In this work, we adopt the data selection, foreground
modeling, and regularization conditions as in \run5 of \citet{Storm:2017arh},
unless stated otherwise.  We perform fits in a region of interest (ROI) of
$|\ell| \leq 90^\circ$ and $|b|\leq 20.25^\circ$, which is important for
component separation, but we restrict most plots to $|\ell| \leq 20^\circ$ and
$|b|\leq 20^\circ$ to highlight the region of the GCE.

\begin{figure*}[t]
  \centering
  \includegraphics[width=0.95\textwidth]{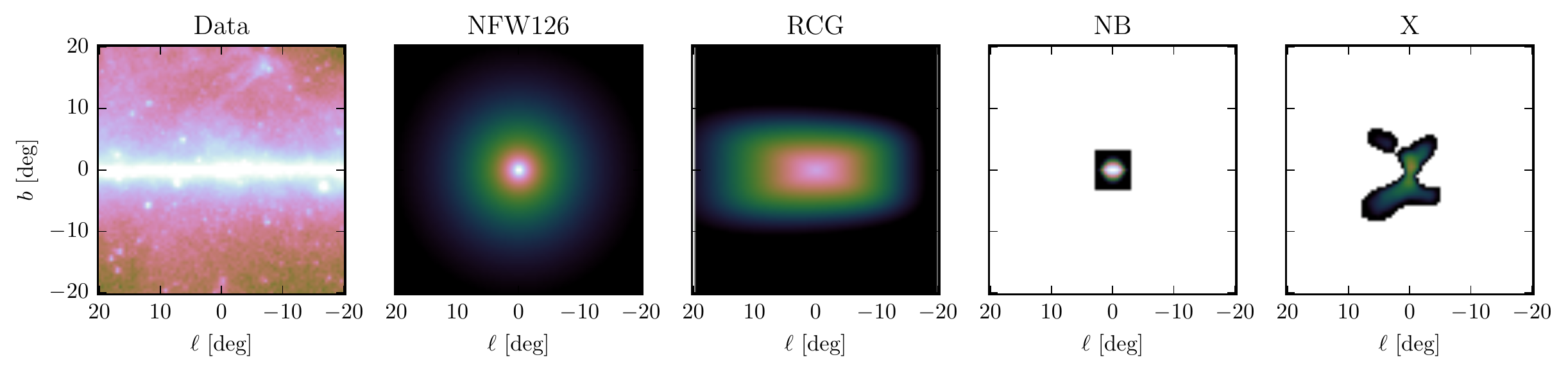}
  \caption{\emph{Left panel:} \Fermi-LAT data above 1 GeV in the inner
    $40^{\circ}\times40^{\circ}$ around the Galactic center.  \emph{Other
    panels:} Spatial templates used to fit the GCE, with arbitrary
    normalization. From left to right: DM profile (NFW126), 
    boxy-bulge,
    nuclear bulge, X-shaped bulge.}
  \label{fig:templates}
\end{figure*}

In order to study the GCE, we model the GCE with fixed spatial templates and
derive the energy spectrum from a fit to the \gr data.  To this end, the following spatial templates are considered: (i) Templates inspired by annihilating DM: two generalized NFW templates with inner slopes of $\gamma=1$ and $\gamma=1.26$ (\texttt{r5\_NFW100} and \texttt{r5\_NFW126}) respectively and an Einasto profile  with $\alpha=0.17$ (\texttt{r5\_Einasto}) \citep{Navarro:1996gj, 1965TrAlm...5...87E, Graham:2005xx, Navarro:2008kc}; (ii) A superposition of two Gaussians and a Galactic central source used to model the 511 keV emission from the inner Galaxy by \cite{Siegert:2015knp}, \texttt{r5\_BulgeGC}; (iii) Templates based on the stellar mass distribution in the Galactic bulge. We adopt a model for the boxy bulge derived from observations of red-clump giants, RCG~\citep{Cao:2013dwa}. We also consider linear combinations of this model with the NB \citep{Launhardt:2002tx} and with the X-shaped bulge \citep{2016AJ....152...14N}.  The addition of the latter is motivated by the recent results from \cite{Macias:2016nev}. These runs are labeled \texttt{r5\_RCG}, \texttt{r5\_RCG\_NB} and \texttt{r5\_RCG\_NB\_X}.  For any linear combination the normalization of each component is left free to vary in each energy bin. Representative examples of these templates, along with the \Fermi-LAT data, are illustrated in Fig.~\ref{fig:templates}. 

All of the runs above are also performed with a fixed MSP-like spectrum instead
of a free spectrum (labeled with the suffix \texttt{\_msp}). For these runs, we use
the stacked MSP spectrum from \cite{McCann:2014dea}, $dN/dE \propto
E^{-1.46}\exp{\left(-E/3.6\right)}$.  Fluxes and significances are derived
using the runs with fixed spectra. 

We emphasize that, given the large modeling uncertainties of cosmic-ray induced
\gr\ emission from the inner Galaxy, we do not explicitly include a
source of cosmic rays at the GC when modeling the diffuse components.
However, such sources are expected, \textit{e.g.}, from star formation in the
central molecular zone (CMZ, \citealt{Gaggero:2015nsa,
Carlson:2016iis, Carlson:2015ona}). The associated emission will depend on the efficiency of cosmic-ray
acceleration, the effects of potentially strong advective winds or anisotropic
diffusion, which are difficult to model in detail.  In our analysis, the expected
hard emission would be instead absorbed by our \Fermi\ Bubbles component (see
supplemental material, \ref{sec:bubbles}, for a discussion).


\section{Results and discussions}
\label{sec:results}

\subsection{Comparison of templates}

\begin{table}[h]
    \centering
    \begin{tabular}{lcc}
      \toprule
      Run & \multicolumn{2}{c}{$-2\ln\mathcal{L}$} \\
                    & free spectrum & MSP spectrum \\
      \midrule
        \texttt{r5\_RCG\_NB\_X} & 647808.1 & 648020.2\\
           \texttt{r5\_RCG\_NB} & 647831.2 & 648027.5\\
               \texttt{r5\_RCG} & 647884.7 & 648061.7\\
           \texttt{r5\_BulgeGC} & 647916.5 & 648140.3\\
           \texttt{r5\_Einasto} & 647961.4 & 648188.6\\
            \texttt{r5\_NFW126} & 648021.8 & 648242.4\\
            \texttt{r5\_NFW100} & 648049.8 & 648278.6\\
      \bottomrule
      \end{tabular}
    \caption{Log-likelihood values for fits with various GCE templates.  Column 2 shows results for a unconstrained GCE spectrum, and column 3 for a spectrum fixed to stacked MSPs.}
\label{tab:GCE_templates}
\end{table}

In Tab.~\ref{tab:GCE_templates} we compare the values of the total (Poisson
plus constraints; see \cite{Storm:2017arh} for details) log-likelihood,
$-2\ln\mathcal{L}$, from the \skyfact\ runs, of the various modifications of
\run5 with different GCE templates with constrained morphology.  We find that,
formally, the combination of boxy bulge as traced by RCG and NB
(\texttt{r5\_RCG\_NB}) provides a better fit to the data than the other runs
(except the one including the X-shaped bulge, see below).  The total flux
associated with the bulge is $\left(2.1\pm 0.1\right)\times
10^{-9}\mathrm{\,erg\,cm^{-2}\,s^{-1}}$ for
the component traced by RCG and $\left(2.3\pm 0.4\right)\times 10^{-10}
\mathrm{\,erg\,cm^{-2}\,s^{-1}}$ for the NB component (in the range $0.1\text{--}100\mathrm{\,GeV}$).  The quoted errors are
statistical; we emphasize that typical systematic uncertainties from modeling assumptions (the range of allowed modulation parameters, etc.) are generally smaller than a factor $\sim 2$.

We find that the addition of the X-shaped bulge can only mildly improve the fit
quality.  Its total flux 
is $(3\pm1)\%$ of that of the boxy bulge for the fixed
spectrum run (\texttt{r5\_RCG\_NB\_X\_msp}). This value is only slightly
smaller than the expectations from \cite{Li:2012qr} and \cite{Cao:2013dwa}, who
find the X-shape to be, by mass, about 6--7\% of the boxy bulge (although
fractions of $20$--$30\%$ \citep{2015MNRAS.448..713P} and $\sim 45\%$
\citep{2015MNRAS.450L..66P} have also been argued).  We find that this
component is not critical for providing a good fit to the data ($2.7\sigma$
improvement), and will concentrate subsequently on the RCG+NB model.  For a
more detailed discussion of the X-shaped bulge and the from
\cite{Macias:2016nev} see the supplementary material~\ref{sec:ROI}.

We find that RCG+NB model provides a significantly better fit than any of the DM
models. These DM profiles can be excluded with a high significance of about
$12.5\sigma$.

\medskip

\begin{figure*}[t]
  \centering
  \includegraphics[width=0.45\linewidth]{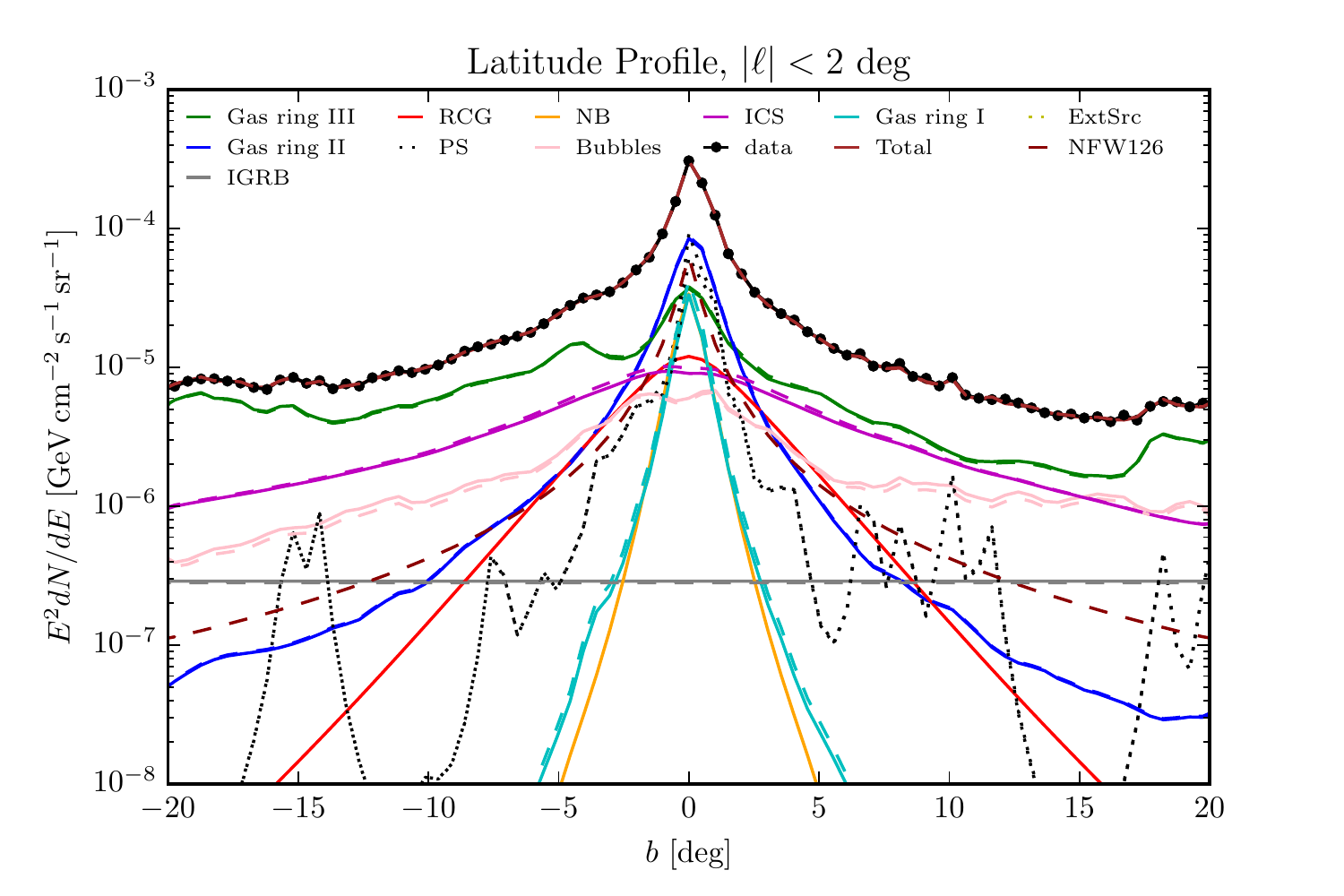}
  \includegraphics[width=0.45\linewidth]{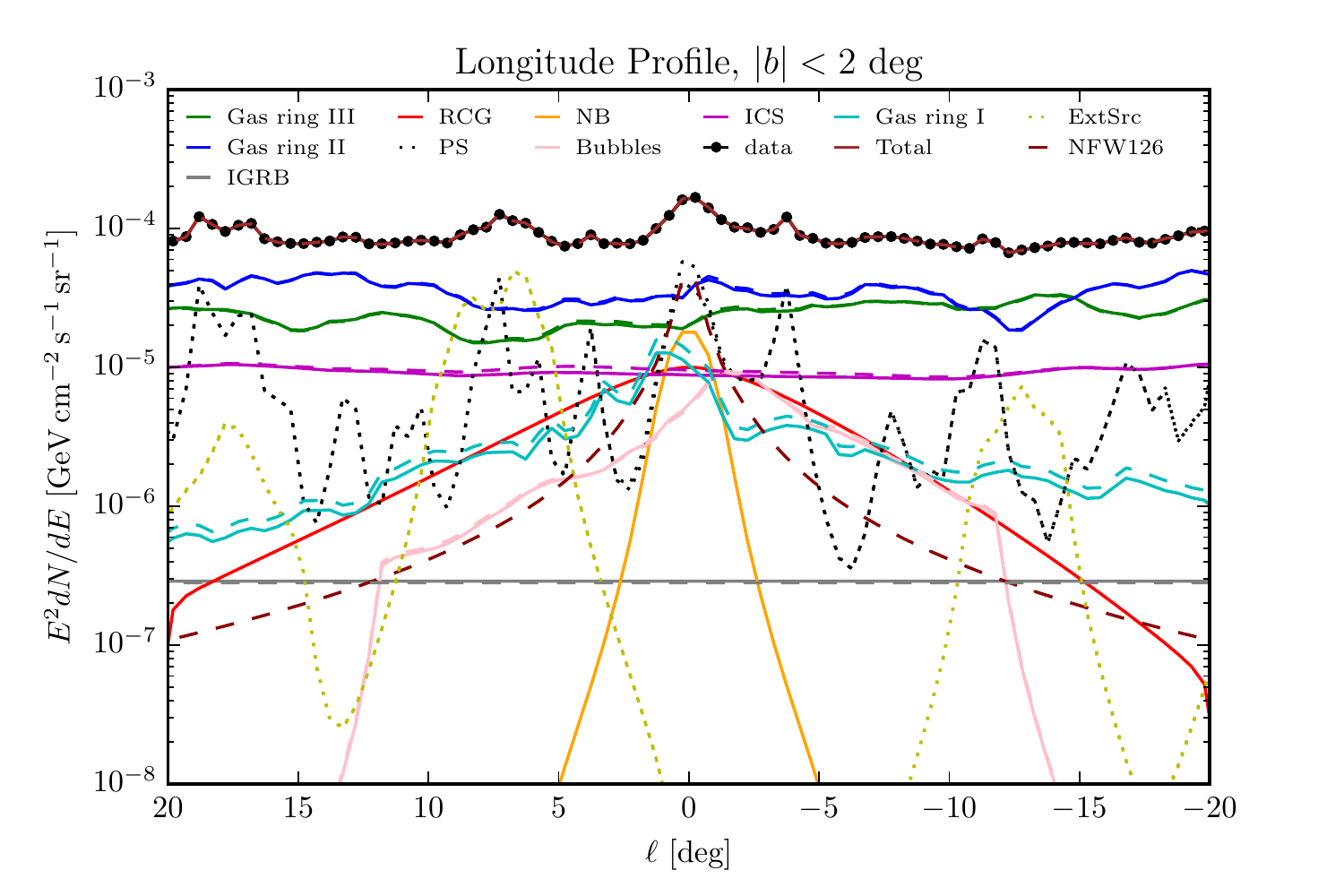}
  \caption{ Measured flux compared to modeled flux as function of Galactic
  latitude (left panel, assuming $|\ell|<2^\circ$) and longitude (right panel,
  assuming $|b|<2^\circ$), for the best-fit run \texttt{r5\_RCG\_NB} (solid lines).  The
  dashed lines show the best-fit fluxes obtained for
  run \texttt{r5\_NFW126} for comparison. Dotted black and yellow lines represent the total point and extended source emission, respectively.}
  \label{fig:profiles}
\end{figure*}

In Fig.~\ref{fig:profiles}, we show the longitudinal and latitudinal
dependences of the various model components compared with \Fermi-LAT data, for
two different GCE models, namely the \texttt{r5\_NFW126} and
\texttt{r5\_RCG\_NB} runs. The solid lines correspond to the components of the
\texttt{r5\_RCG\_NB} run, while the dashed lines of the same color correspond to
the \texttt{r5\_NFW126} components, except for the GCE component, which is red (RCG) and orange (NB) for the \texttt{r5\_RCG\_NB} run and brown (NFW126) for the \texttt{r5\_NFW126} run. The dotted black and yellow lines are point sources and extended sources, respectively, which have the same total flux in both runs. There is very little variation in any
components except those of the GCE (in the latitude profile, the extended
source flux peaks just below the lower limit of the plot). The shape differences between the RCG+NB templates compared
to the NFW template are, however, quite large. The NFW is much more strongly
peaked, and is of course spherically symmetric, while the oblateness and
asymmetry of the RCG profile can be seen by comparing the shape of the tails in
the latitude and the longitude profile plots.

\begin{figure*}[t]
  \centering
  \includegraphics[width=0.45\linewidth]{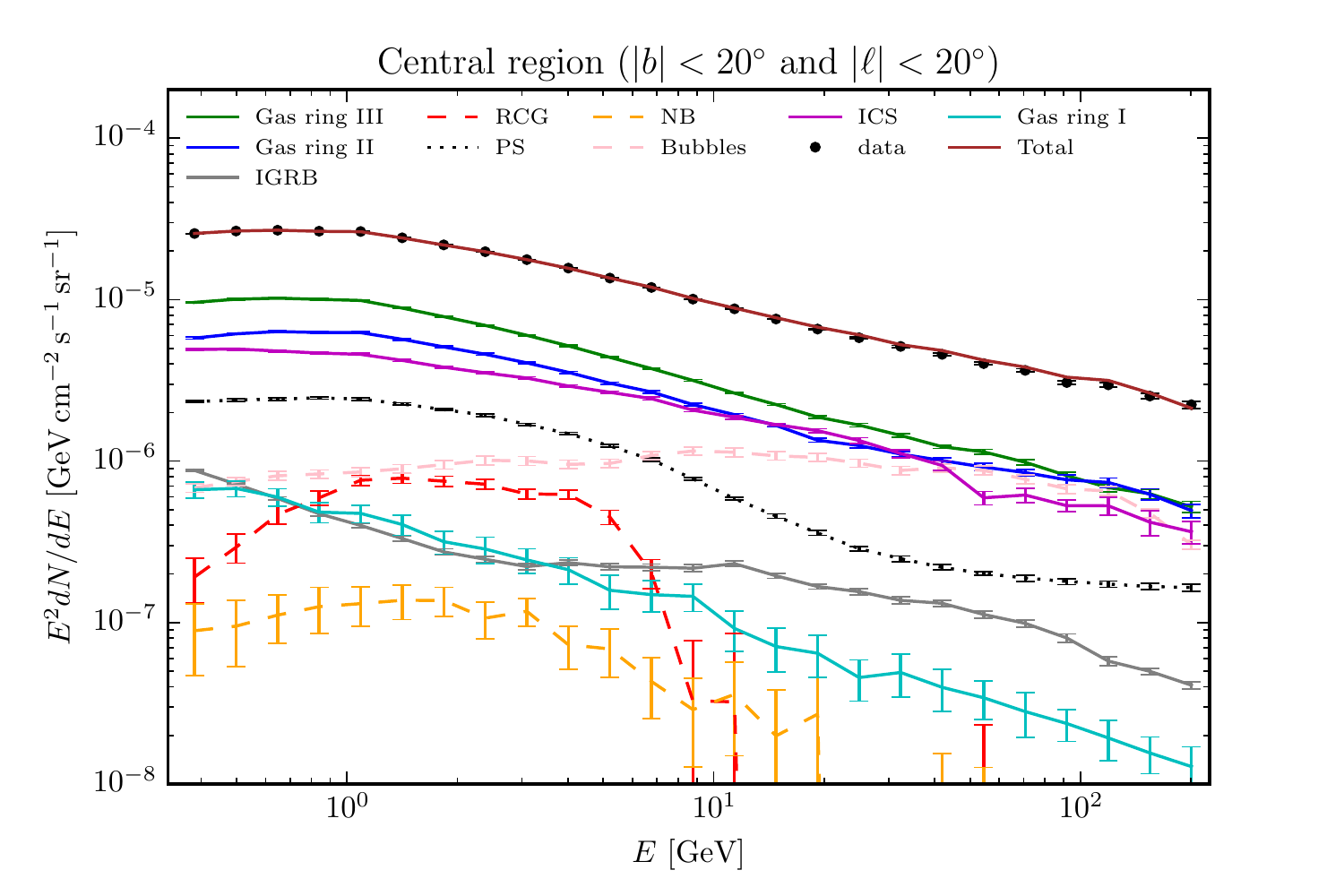}
  \includegraphics[width=0.45\linewidth]{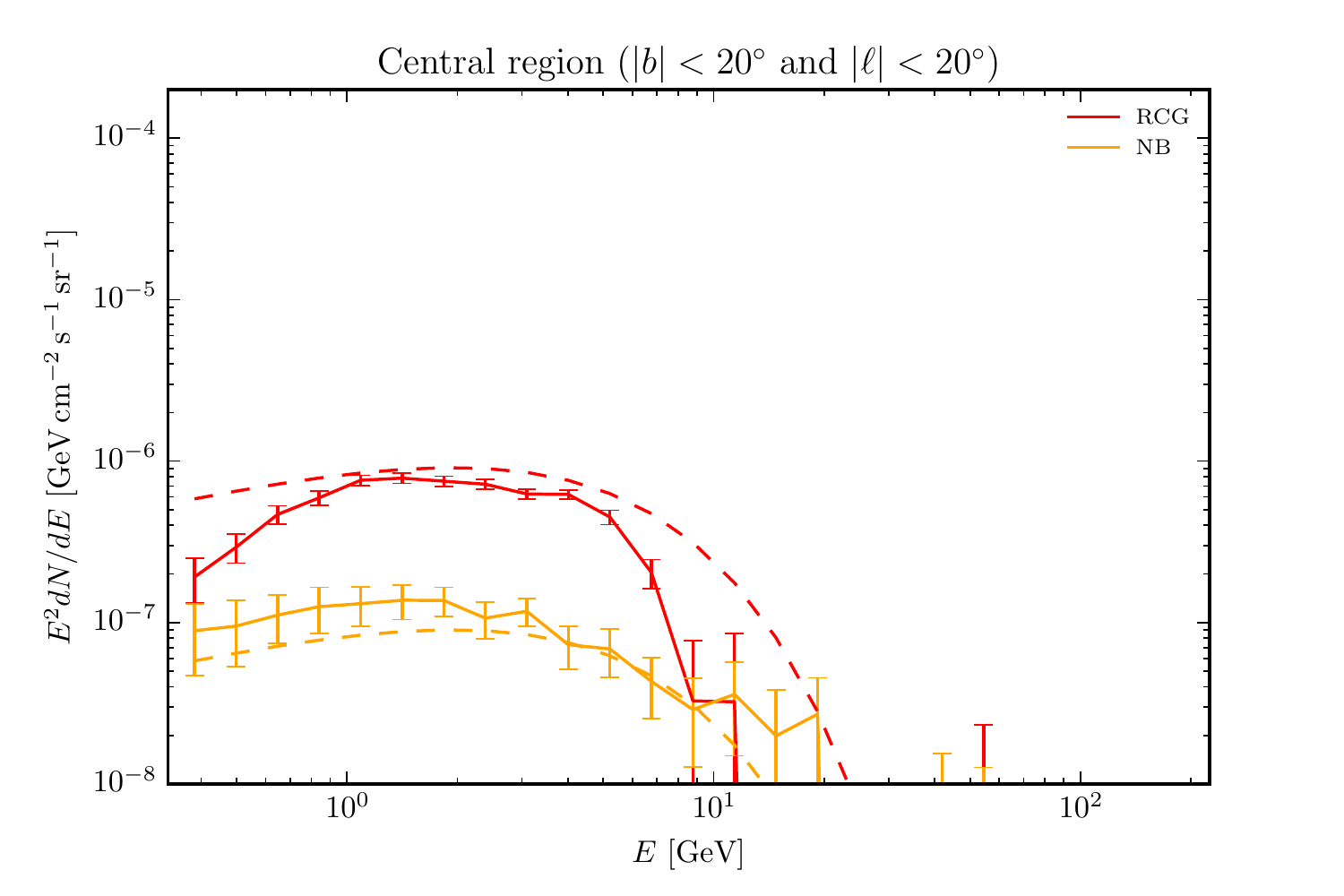}
  \caption{
    \emph{Left panel:} Spectrum of the various model components in the region
    $|\ell|,|b| < 20^\circ$, as function of photon energy, for run
    \texttt{r5\_RCG\_NB}. \emph{Right panel:} GCE spectra of RCG and NB
    component both for free spectra (solid) and for fixed spectra (dashed). 
    }
  \label{fig:GCE_spectra}
\end{figure*}

In Fig.~\ref{fig:GCE_spectra} on the left hand side, we show the spectra for
all components of the \texttt{r5\_RCG\_NB} run in the inner
$40^{\circ}\times40^{\circ}$ around the Galactic center. On the right, we show
the spectra for the \texttt{RCG} and \texttt{NB} components for separate runs
where the spectral shape was left free to vary in one case and fixed to an
MSP-like spectrum (with free overall normalization) in the other in the same
region. The results from the free-spectra and fixed-spectra runs agree
reasonably well, although the spectrum of the RCG component is somewhat more
pronounced in the free-spectrum run.  However, we find that the general
preference for the RCG+NB scenario over DM-inspired templates is the same in
both cases.

\subsection{Light/mass ratios}

We now estimate the light-to-mass ratio for the RCG and NB components
separately.  The stellar mass of the nuclear bulge is $(1.4\pm
0.6)\times10^9\mathrm{\,M_\odot}$ \citep{Launhardt:2002tx}, while the mass of
the boxy-bulge is $(0.91\pm 0.7)\times10^{10}\mathrm{\,M_\odot}$
\citep{Licquia:2014rsa}\footnote{ The bulge mass from \cite{Licquia:2014rsa} is
derived by combining bulge mass estimates from the literature in a hierarchical
Bayesian analysis. We take this bulge mass as our reference value.  However, we
note that individual estimates range in best-fit value from $0.48\times
10^{10}\mathrm{\,M_\odot}$ to $2.74\times 10^{10}\mathrm{\,M_\odot}$, the range
coming from different model assumptions and measurement techniques
\citep[see][for a thorough discussion]{Licquia:2014rsa}.}.

Combining the mass measurements with the luminosities of the boxy-bulge and
nuclear bulge components (mentioned above), the light-to-mass ratio for the
bulge component is found to be $(2.1\pm0.2)\times
10^{27}\mathrm{\,erg\,s^{-1}\,M_\odot^{-1}}$ , and for the NB component
$(1.4\pm0.6)\times 10^{27}\mathrm{\,erg\,s^{-1}\,M_\odot^{-1}}$, from
$0.1\text{--}100\mathrm{\,GeV}$.  The light/mass ratios of the two components
are consistent within uncertainties, providing further circumstantial evidence
that the GCE emission is correlated with stellar mass in the bulge.

\begin{figure}[t]
  \begin{center}
  \includegraphics[width=0.95\linewidth]{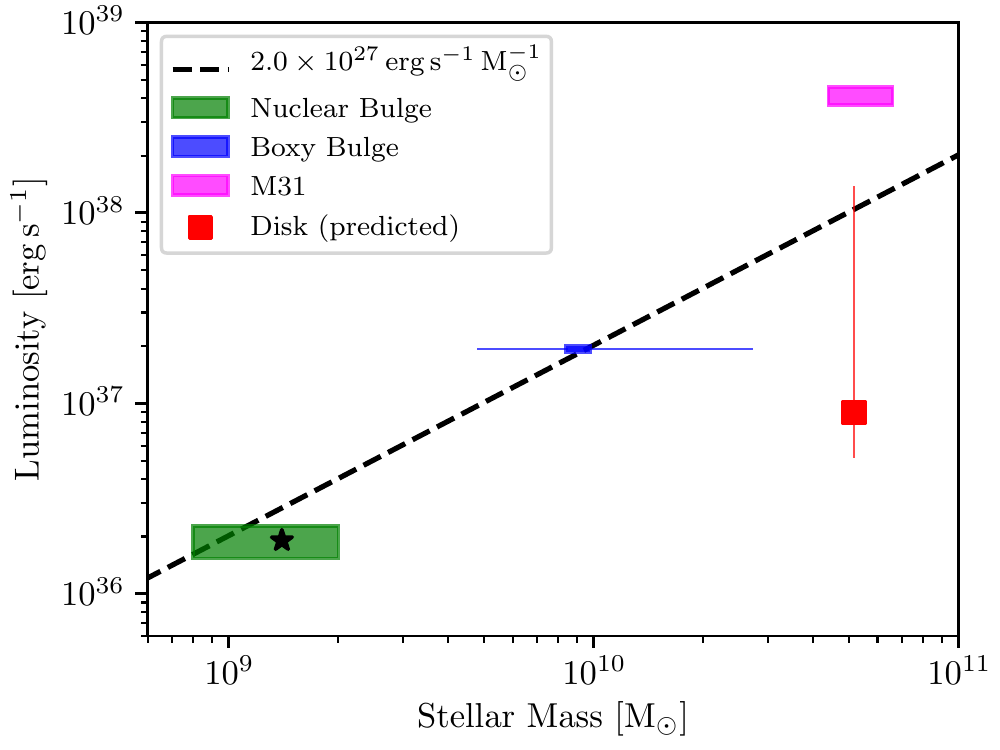}
  \end{center}
  \caption{Stellar mass compared to the observed \gr\ luminosity for the boxy
    bulge (blue) and the nuclear bulge (green). Widths correspond to the
    uncertainty in the mass estimates, with the star indicating the best-fit
    mass of the NB \citep{Launhardt:2002tx}. The boxy bulge mass comes is
    derived from a compilation of measurements into a hierarchical Bayesian
    analysis (box) the thin line displays the range of individual measurements
    \citep{Licquia:2014rsa}.  Heights reflect the uncertainty in the derived
    GCE flux.  We also show the emission from M31 \citep{Ackermann:2017nya}
    assuming that all \gr\ emission comes from its bulge, see methods
    Sect.~\ref{sec:M31} for further details.  The long-dashed line gives the
    relation of GCE emission per unit of stellar mass that best fits the
    combination of boxy bulge and nuclear bulge.
    Finally, the expected disk luminosity is shown as the red box.
    The estimate suffers from a large uncertainty due to the unknown
    \gr\~luminosity function, this uncertainty is bracketed by the thin red
    line.
    }
  \label{fig:mass2gev}
\end{figure}

The relation between stellar mass and the GCE luminosity is illustrated in
Fig.~\ref{fig:mass2gev}.  It shows the observed GCE intensity of various
components compared to their stellar mass.  This figure shows that, within
uncertainties, the GCE emission indeed scales with stellar mass of the RCG and
the NB component.  Also shown is the excess of $\gamma$ rays recently
observed from the direction of M31, interpreted as a potential ``GCE" in this galaxy \citep{Ackermann:2017nya}. We
find that if this interpretation were correct, it would correspond to a larger emission per unit stellar mass by a factor $\sim 4$ than what is observed in the Milky Way (for details see the Methods section \ref{sec:M31}).

Given that MSPs are the most likely candidate source class for the GCE, it is
useful to quantify the corresponding emission expected from MSPs in the Galactic
disk. We estimate the flux from the
MSP disk population, using 3FGL \gr\ flux
measurements of local MSPs~\citep{Acero:2015hja} and distance and period
information from the ATNF catalog \citep{Manchester:2004bp} (for details see
supplementary material, Sect.~\ref{app:BtoD}).  From this, the
expected bulge-to-disk flux (luminosity) ratio is $\sim 0.9\,(2.3)$ which
implies a $\sim 10\times$ larger number of MSPs per unit of stellar mass in the
bulge compared to the disk.  Interestingly, this number is comparable to what is
measured for another mysterious emission in the inner Galaxy, namely the 511
keV positron-annihilation-line emission~\citep{Knodlseder:2005yq}.  This
so-called 511 keV line emission has also been observed in the disk, with the
latest estimate for the bulge-to-disk flux ratio being $B/D=0.58\pm0.13$
\citep{Siegert:2015knp}. 
We stress however that the estimate for the MSP bulge-to-disk ratio
is highly dependent on the assumed high-flux slope of the MSP luminosity function, 
which is quite unconstrained~\citep{Strong:2006hf,Venter:2014zea,Calore:2014oga,Winter:2016wmy, Eckner:ToAppear}.
See Methods section \ref{app:BtoD} for a more detailed discussion.

For the above estimate, the expected emission from MSPs in the disk is $\sim8\%$ of the ICS flux
and $\sim 2\%$ of the $\pi^0$ flux (in the 1-10 GeV band). Since the unresolved part of the disk MSP population
makes up an even smaller fraction, we expect the corresponding diffuse contribution to be
undetectable above the foregrounds and backgrounds.

\subsection{Comparison with previous results}

Previous analyses have typically found that the morphology of the GCE is spherically symmetric ~\citep{Daylan:2014rsa, Calore:2014xka}, although this is challenging to prove decisively mainly because of the contamination of the \Fermi~bubbles at low latitudes~\citep{TheFermi-LAT:2017vmf}. Additionally, some degree of
elongation was found at high energies~\citep{Linden:2016rcf}.  Our findings are significantly different in this respect.  We find that the critical difference between
the previous and the current analyses is the inclusion of modulation parameters
that account for uncertainties in the gas and ICS templates.  The magnitude of
these variations is completely expected, given the large uncertainties of the
various templates (see the supplementary material \ref{sec:template2skyfact} for a
detailed comparison).  Our results are stable against variations of the
regularization parameters (see supplementary material \ref{sec:mod} for a
discussion).

\section{Conclusion}
\label{sec:conclusions}

We presented a novel analysis of the Galactic center GeV excess using the
recently developed tool \skyfact\ \citep{Storm:2017arh}, allowing for the
inclusion of a large number of nuisance parameters to take into account
uncertainties in the foreground models.  The effects of star formation in the
CMZ are not directly modeled, but associated hard emission is captured in the
low-latitude part of our \Fermi\ Bubbles component.  We studied in detail the
morphology of the GCE and we compared it against various templates, most
notably the contracted NFW profile previously shown to describe the excess and
a model that traces the stellar distribution of the bar/boxy bulge and nuclear
bulge in the inner Galaxy.  

We demonstrated that the stellar bulge model provides a significantly better
fit ($>10\sigma$) to the data than the DM-emission related Einasto or
contracted NFW profiles.  Hence the GCE appears to simply trace stellar mass in
the bulge, not the dark matter density squared (although the actual DM profile
is sufficiently uncertain that this possibility cannot be entirely excluded).
What is more striking is that the light/mass ratio that we independently derive for
the boxy bulge and the nuclear bulge are consistent with each other, supporting
this interpretation. 

The arguably best candidate sources are MSPs in the Galactic
bulge.  
The putative bulge MSP population can
be efficiently probed in the upcoming years with searches for radio pulsation
signals~\citep{Calore:2015bsx} with MeerKAT or SKA.  Our findings provide
important information to guide these searches and strong motivation to perform
them vigorously.


\acknowledgments

\paragraph{Acknowledgments} We thank Lia Athanassoula, 
David Berge, Gianfranco Bertone, Ilias
Cholis, Roland Crocker, Oscar Macias, Pasquale Serpico, Tracy Slatyer, Andy Strong, Jacco Vink
and Gabrijela Zaharijas for useful discussion. 
We acknowledge Daniele Gaggero
for the support provided with the DRAGON code.  RB would like to thank the
organizers and participants of the TeVPA 2017 mini-workshop on the GCE for a
fruitful discussion.  This research is funded by NWO through the
Vidi research program "Probing the Genesis of Dark Matter" (680-47-532; ES and CW), and
through a GRAPPA-PhD fellowship (RB).
FC acknowledges support from Agence Nationale de la Recherche under the contract ANR-15-IDEX-02, 
project "Unveiling the Galactic centre mistery", GCEM (PI: F. Calore).


\bibliographystyle{aasjournal}
\bibliography{GCE_stars}

\newpage
\appendix

\section{Methods}
\label{sec:methods}

\subsection{GCE templates}
\subsubsection{Template construction}

We construct various spatial models to study the morphology of the GCE. Below
we provide some additional motivation for the various templates and describe
their construction.  

When we construct the template from the density profile, we perform a
line-of-sight (l.o.s.) integral over the density profile.
\begin{equation}
  \label{eq:los}
  \frac{d\Phi}{d\Omega}\propto
  \left\{
  \begin{split}
    & \int_\mathrm{l.o.s.} ds\,\rho^2(r(s)) ds \qquad \text{(DM)}\\
    & \int_\mathrm{l.o.s.} ds\,\rho(r(s)) ds \qquad \text{(Stars)}
  \end{split}
  \right.\;.
\end{equation}
Here the parameter $s$ describes the los, $r(s)$ is the distance away from the
Galactic center, $\rho$ is the density, $\Phi$ the flux on earth and $\Omega$
the solid angle.  For annihilating DM the emission traces the density squared.
Furthermore, we define the following coordinates centered on the GC:
\begin{align}
	x_\mathrm{GC}(s, \ell, b) &= R_\mathrm{\odot} - 
    	s \cos\left(b\right) \cos\left(\ell\right), \\
    y_\mathrm{GC}(s, \ell, b) &= s \cos\left(b\right) \sin\left(\ell\right), \\
    z_\mathrm{GC}(s, b) &= s \sin\left(b\right),
\end{align}
where $R_\odot = 8.3\mathrm{\,kpc}$ is the distance from the Sun to the GC
\citep{Gillessen:2008qv}, $\ell$ is Galactic longitude and $b$ Galactic
latitude. Note that the solar system lies along the x-axis.  The Galactocentric
radius is now given by $r=\sqrt{x_\mathrm{GC}^2 + y_\mathrm{GC}^2 +
z_\mathrm{GC}^2}$.

\subsubsection{Templates considered}

\paragraph{Positron annihilation signal / $511 \mathrm{\,keV}$ line} 
The all-sky positron-annihilation signal has a strong component corresponding
to the Galactic bulge \citep{Knodlseder:2005yq}.  Galactic positron emission
can be described by a disk component and a bulge component.  The latter is well
described by a superposition of two Gaussians, the broad and narrow bulge
components \citep{Bouchet:2010dj, Siegert:2015knp}. In addition,
\cite{Siegert:2015knp} finds evidence for the presence of a central component
that is consistent with being a point source.  We model the 511 keV bulge using
the spatial profile and intensity as described in Tables 2 and 6 of
\cite{Siegert:2015knp}. We include the central source component, the narrow and
broad bulge and refer to this as \texttt{r5\_BulgeGC}. For the bulge we assume
that the narrow (broad) component contains 28\% (72\%) of the total bulge flux
\citep{PoS(Integral2014)054}.

\paragraph{Boxy bulge}
We model the boxy, or pseudo, bulge using the distribution of
RCGs \citep{Nataf:2012pk,Cao:2013dwa}.  The number density of RCGs is well
fit by the triaxial \textit{E3} model \citep{Dwek:1995xu,Cao:2013dwa}:

%
\begin{align}
  n_\mathrm{E3} &\propto K_0(r_s) \\
  r_s &= \left[\left[
  \left(\frac{x}{x_0}\right)^2 + 
  \left(\frac{y}{y_0}\right)^2\right]^2
  + \left(\frac{z}{z_0}\right)^4
  \right]^\frac{1}{4},
\end{align}
with $K_0$ the modified Bessel function of the second kind and
$x_0=0.67\mathrm{\,kpc}, y_0=0.29\mathrm{\,kpc}$ and $z_0=0.27\mathrm{\,kpc}$
scale lengths.  We use the best-fit parameters from \cite{Cao:2013dwa} for the
E3 model. A larger set of parametric models to describe the triaxial structure
of the bulge are given in \cite{Dwek:1995xu}.

In case of the boxy bulge it is important that it is rotated.  In order to
perform the line-of-sight integral we have to perform coordinate transformation
$(x,y,z)\rightarrow(x_\mathrm{GC}, y_\mathrm{GC}, z_\mathrm{GC})$.  The
rotation of the major axis ($x$) is around the $z$--axis and is
$\theta=29.4^\circ$ away from $y_\mathrm{GC}$ in the counterclockwise direction
\citep{Cao:2013dwa}.  So the conversion is given by $x =
y_\mathrm{GC}\cos\theta + x_\mathrm{GC}\sin\theta$ and $y =
x_\mathrm{GC}\cos\theta - y_\mathrm{GC}\sin\theta$.

\paragraph{Nuclear bulge}
The Nuclear Bulge (NB) is a distinct component in the inner part of our
$\lesssim 300\mathrm{\,pc}$ of our Galaxy with ongoing star formation
\citep{1996Natur.382..602S}.  It consists out of two components, the nuclear
stellar disk (NSD) and the nuclear stellar cluster (NSC)
\citep{Launhardt:2002tx}.  We model these two components following
\cite{Launhardt:2002tx}, who find a total mass for the NB of
$(1.4\pm 0.6) \times 10^9\mathrm{\,M_\odot}$. For the mass density of the NSC
we use
\begin{equation}
  \rho(r)=\begin{cases}
    \frac{\rho_{0, \mathrm{NSC}}}{1+\left(\frac{r}{r_0}\right)^2}, & \text{$r\leq 6\mathrm{\,pc}$} \\
    \frac{\rho_{1, \mathrm{NSC}}}{1+\left(\frac{r}{r_0}\right)^3}, & \text{$6\mathrm{\,pc}<r\leq 200\mathrm{\,pc}$} \\
    0, & \text{$r>200\mathrm{\,pc}$}.
  \end{cases}
\end{equation}
Here $\rho_{0, \mathrm{NSC}} =  3.3\times 10^6\mathrm{\,M_\odot\,pc^{-3}}$ and
$\rho_{1, \mathrm{NSC}}$ is defined such that the profile is continuous at the
break.

The NSD is modeled as a combination of an exponential disk with a scale height
of $45\mathrm{\,pc}$ and a broken power law for the radius:
\begin{equation}
  \rho(r)=\begin{cases}
    \rho_{0, \mathrm{NSD}} \left(\frac{r}{1\mathrm{\,pc}}\right)^{-0.1}
    e^{-\frac{\left|z\right|}{45\mathrm{\,pc}}}, & \text{$r < 120\mathrm{\,pc}$} \\
    \rho_{1, \mathrm{NSD}} \left(\frac{r}{1\mathrm{\,pc}}\right)^{-3.5} 
    e^{-\frac{\left|z\right|}{45\mathrm{\,pc}}}, &\text{$120\mathrm{\,pc}\leq r < 220\mathrm{\,pc}$} \\
    \rho_{2, \mathrm{NSD}} \left(\frac{r}{1\mathrm{\,pc}}\right)^{-10} 
    e^{-\frac{\left|z\right|}{45\mathrm{\,pc}}}, &\text{$r\geq 220\mathrm{\,pc}$}.
  \end{cases}
\end{equation}
With $\rho_{0, \mathrm{NSD}} = 301 \mathrm{\,M_\odot\, pc^{-3}}$ such that the
mass within $120\mathrm{\,pc}$ is $8\times10^{8}\mathrm{\,M_\odot}$.  Again,
$\rho_{1, \mathrm{NSD}}$ and  $\rho_{2, \mathrm{NSD}}$ are defined such that
the density profile is continuous. 

\paragraph{X-shape}
An X-shaped component is seen in the distribution of RCGs in
the boxy bulge \citep{McWilliam:2010bc, Nataf:2010wf, 2016AJ....152...14N}.
Such an X-shaped structure is a generic feature of pseudo bulges, believed to
arise through the buckling instability \citep{Li:2012qr}.

We model the X-shaped bulge using the results from \cite{2016AJ....152...14N}
and the public release of the WISE coadds \citep{2014AJ....147..108L}.  We take
the average of the residual maps in the W1 and W2 bands obtained using the
public code\footnote{Code available at \texttt{http://unwise.me}} from
\cite{2016AJ....152...14N}, revealing the X-shaped structure.  Any negative
residuals are set to zero before averaging. This procedure is similar to that
adopted by \cite{Macias:2016nev}.

\paragraph{DM profiles}
We test two different classes of DM density profiles.  First, we consider the
NFW \citep{Navarro:1996gj, Graham:2005xx} profile:
\begin{align}
  \rho(r) = \frac{\rho_s}{
  \left(\frac{r}{r_s}\right)^\gamma
  \left(1 + \frac{r}{r_s}\right)^{3-\gamma}
  }.
\end{align}
here $\gamma$ defines the inner-slope, $r_s=20\mathrm{\,kpc}$ the scale-radius
and $\rho_s$ the scale density.  We consider the common NFW profile with
$\gamma=1$ and a contracted NFW profile with $\gamma=1.26$.

The second profile considered is the Einasto profile
\citep{1965TrAlm...5...87E, Navarro:2008kc}:
\begin{align}
  \rho(r) = \rho_s
  \exp\left(-
  \frac 2 \alpha\left[
  \left(\frac{r}{r_s}\right)^\alpha - 1
  \right]\right).
\end{align}
Where we set $\alpha=0.17$ and again we use $r_s=20\mathrm{\,kpc}$.

Since we are only interested in the morphology the overall normalization is
irrelevant, however, for completeness we mention that the density profile can
be normalized using the DM density in the solar neighbourhood
$\rho\left(R_{\odot}\right) = 0.4 \mathrm{\,GeV\,cm^{-3}}$
\citep{Read:2014qva}. Also note that the exact value of $r_s$ is not very
important, since this mostly affects the halo properties at large radii.

%
\begin{figure}[t]
  \centering
  \includegraphics[width=0.5\linewidth]{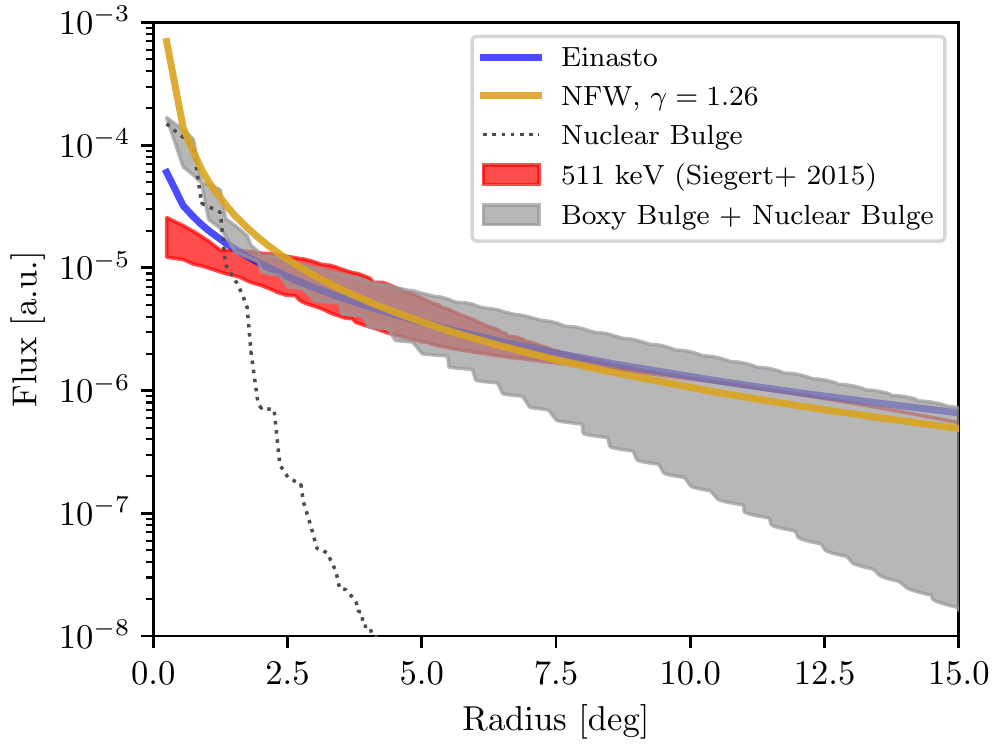}
  \caption{Radial profiles for the various GCE templates considered in this
    analysis.  All templates are smoothed with the \Fermi\ Pass 8
    \texttt{ULTRACLEAN} Front+Back angular resolution averaged between
    $1\text{--}10\mathrm{\,GeV}$.  The relative normalization between the NB
    and boxy bulge is determined by their total stellar masses $1.4\times10^9
    \mathrm{\,M_\odot}$ and $9.1\times10^{10} \mathrm{\,M_\odot}$ respectively.
    Steps in the grey and red bands (line) are the result of the finite pixel
    size.}
  \label{fig:radial_templates}
\end{figure}

\subsubsection{Template comparison}

The radial profiles of the adopted spatial templates are compared in
Fig.~\ref{fig:radial_templates}, where in case of a non-spherical template we
show the \textit{envelope} of the radial profile (for spherical templates, it
is a single line).  Templates are smoothed with the \Fermi\ Pass 8
\texttt{ULTRACLEAN} Front+Back angular resolution averaged between
$1\text{--}10\mathrm{\,GeV}$.  The spherically-symmetric DM profiles strongly
peak towards the Galactic center.  The 511 keV line emission model is less
peaked towards the center and has a broader distribution, due to the small
offset of one of the model components towards negative longitudes
\citep{Siegert:2015knp}.  The combination of boxy bulge and nuclear bulge,
weighted by their total stellar mass \citep[$1.4\times10^{9}\mathrm{\,M_\odot}$
for the nuclear bulge and $0.91\times10^{10}\mathrm{\,M_\odot}$ for the boxy
bulge][]{Launhardt:2002tx, Licquia:2014rsa} is also highly peaked towards the
center because of the radial distribution of stars in the nuclear bulge.  The
stellar template follows the NFW profile out to almost $\sim 10^\circ$, but it
has a much wider spread at larger radii due to its oblateness. 


\subsection{Analysis with \skyfact}

As mentioned in the main text, we use \skyfact\ for fitting the \gr\ sky. We
refer the reader to \citet{Storm:2017arh} for a complete description of the
approach. Here we briefly describe the approach to fitting, components used in
modeling, and statistical analysis used in \skyfact, highlighting any
differences between this analysis and that in \citet{Storm:2017arh}.

\subsubsection{Fits with \skyfact}

The main difference between fits with \skyfact\ and more traditional template
fitting is that \skyfact\ allows for the introduction of nuisance parameters
that can account for small, pixel-by-pixel variations and uncertainties in the
spatial templates used in fitting. We model the diffuse flux in pixel $p$ and
energy bin $b$ as the sum over $k$ emission components,
\begin{equation}
    \phi_{pb} = \sum_{k}
    T_{p}^{(k)} \tau_{p}^{(k)} \cdot
    S_{b}^{(k)} \sigma_b^{(k)} \cdot
    \nu^{(k)}\;,
    \label{eqn:diffmodel}
\end{equation}
where $T_p^{(k)}$ describes the morphology of emission component $k$,
$S_b^{(k)}$ the model spectrum, and $\nu^{(k)}$ is an overall normalization.
The parameters $\tau_{p}^{(k)}$ and $\sigma_b^{(k)}$ are, respectively, spatial
and spectral modulation (or nuisance) parameters that account for uncertainties
in the model morphology and model spectrum. These parameters are meant to vary
around values of one, in a range that corresponds to our actual knowledge of
the spectrum or morphology of each emission component. They are constrained to be non-negative to remain physically meaningful. The point source model is similar in
structure, but without spatial templates or modulation parameters; see
\citet{Storm:2017arh} for the full model description.

We adopt a penalized likelihood approach, where the flexibility of the
modulation parameters is controlled via regularization (or penalty) terms in
the total likelihood function in the fit. The analytical structure of our
regularization terms is motivated by the maximum-entropy method (MEM), as
discussed in~\cite{Storm:2017arh}.  The diffuse flux is then multiplied by the
exposure and convolved with the \Fermi\ point spread function (PSF) to obtain
the expected number of photons per pixel and energy bin.  The predicted number
of photons per bin are finally compared to the Fermi data using a Poisson
likelihood.

Although the number of parameters in the above minimization (of
$-2\ln\mathcal{L}$) problem is very large, typically of the order of $10^5$,
the minimization problem is, for reasonably constrained modulation parameters,
convex and hence has a unique solution~\citep{Storm:2017arh}.  We use the
L-BFGS-B~\citep{Byrd:1995a,Zhu:1997a,Morales:2011a} algorithm for this purpose,
which makes use of analytical gradient information and implements the active
set method to enforce non-negative boundaries on the fitting parameters.  We
estimate the uncertainties on fitted model parameters using the Hessian of the
likelihood function, which we approximate by the inverse Fisher information
matrix (as described in detail in \cite{Storm:2017arh}; see also
\cite{Edwards:2017mnf}).

\subsubsection{Data selection and model components}

Data selection is identical to \citet{Storm:2017arh}: we use 7.6 years of Pass
8 \texttt{ULTRACLEAN} \Fermi\ data binned into square pixels $0.5^{\circ}$ on a side.  Our
foreground model consists of standard Galactic and extragalactic components: 1)
hadronic emission from the $\pi^0$ decay produced by the interactions of cosmic
ray protons with Galactic gas and dust, 2) ICS emission from cosmic ray
electrons, 3) the extragalactic Isotropic Gamma-Ray Background (IGRB) emission,
primarily the result of unresolved point sources, and 4) point sources and
extended emission sources. We include an additional component that represents
the \Fermi\ Bubbles \citep{Fermi-LAT:2014sfa}. Instead of the $511$~keV
template used for the GCE component in \citet{Storm:2017arh}, we here set up a
variety of runs with different spatial templates for the GCE, detailed in
Section~\ref{sec:modeling}.

As a tracer for the spatial distribution of $\pi^0$ decay \gr\, emission, we
consider the sum of the atomic and molecular hydrogen column densities,
assuming a constant conversion $X_{CO} = 1.9 \times 10^{20}$ cm$^{-2}$/(K
cm/s). We build maps from those available in the GALPROP public
release\footnote{\url{https://galprop.stanford.edu/}}. We split the gas
templates in three radial bins: 0 -- 3.5 kpc, 3.5 -- 6.5 kpc and 6.5 -- 19 kpc.
For the three gas templates we impose a weak regularization on the morphologies
(corresponding to variations within 32\% for in the inner two rings and 50\%
for the outer ring), a weak regularization on the spectra starting from the
ones measured in~\cite{2012ApJ...750....3A} (within 25\%), and a very weak
smoothing. The normalization is instead completely unconstrained.  

To compute the ICS template, we use the public code
DRAGON~\citep{2008JCAP...10..018E} and its custom companion
GammaSKY~\citep{2012PhRvL.108u1102E,2013JCAP...03..036D}. The ICS \gr emission model
is built under standard assumption on the cosmic-ray source
distribution~\citep{2001RvMP...73.1031F} and interstellar radiation
field~\citep{2005ICRC....4...77P}, while we use propagation parameters
corresponding to the ``KRA4'' model, see~\cite{2013JCAP...03..036D}.  The ICS
morphology is allowed to strongly vary (by a factor of 3).  The ICS spectrum,
as in the case of the hadronic template, is constrained to remain close to the
spectrum measured in~\cite{2012ApJ...750....3A} (within $\sim$ 25\%). We do
enforce a stronger smoothing than in the case of the gas, about 10\% pixel-to-pixel variation. The overall normalization is again free to float.
 
We further include a template for the IGRB as
measured in~\citet{2015ApJ...799...86A}.  We allow its spectrum to slightly
vary (within 25\%) with respect to the spectrum derived from the diffuse
background model A in~\citet{2015ApJ...799...86A}, while its normalization is
fixed.

We add all 3FGL point sources within our ROI to the fit~\citep{Acero:2015hja}.
We use a weak regularization on their spectra, allowing variations within 20\%,
while the overall normalizations are allowed to vary by 32\%.  We also consider
3FGL extended sources~\citep{Acero:2015hja} within our ROI, weakly constraining
the spectra and leaving the morphologies unconstrained (see details of the
implementation in~\citealt{Storm:2017arh}). 

As in \citet{Storm:2017arh}, we use a constrained (to within $5\%$) spectral
template from \citet{Fermi-LAT:2014sfa} with unconstrained spatial modulation
allowed for the \Fermi\ Bubbles component. In the supplementary information
\ref{sec:supp}, we discuss the effects of using alternative templates for this component.

The main technical difference between \run5 in \citet{Storm:2017arh} and this
paper are the models used for the GCE. For each of the tested GCE components
listed in the main text, the spatial modulation is fixed while the spectral
modulation is set to be either essentially free (\texttt{r5\_*}) or completely
fixed to an MSP spectrum (\texttt{r5\_*\_msp}).

Lastly, we tested that a more complex version of the \texttt{r5\_*} runs, where
we split the CO and HI components of the outer ring and include a free
template for the CMZ (which covers only the inner 4 pixels), does not affect the
results significantly.  Finally, we show in the supplementary
material~\ref{sec:supp} that the impact of including the 2FIG catalog
from~\cite{Fermi-LAT:2017yoi} is minimal: the RCG+NB template is still
preferred over the DM templates.

\subsubsection{Statistical Analysis}

The full likelihood used in \skyfact\ is the sum of a standard Poisson
likelihood and a regularization term that control the modulation parameters; a
complete description is available in \citet{Storm:2017arh}. The full likelihood
is used to calculate any significances listed in the paper.

For the comparisons between different GCE templates in
Section~\ref{sec:results} of the main text, we perform a standard likelihood
ratio test using the full likelihood. We use only the runs with the fixed MSP
spectra. For these runs, the difference in the degrees of freedom is simply
equal to the the overall normalizations. When the spectra are left free, the
difference in the degrees of freedom is less clear, due to the regularization.
We describe a method to estimate the \textit{effective} degrees of freedom with
mock data in \citet{Storm:2017arh}, but for clarity, here we choose to compare
the runs with the spectral degrees of freedom fixed (we also did this in
\citealt{Storm:2017arh} to estimate the significance of adding a GCE
component). When comparing \texttt{r5\_RCG\_NB} and \texttt{r5\_Einasto}, the
difference in the degrees of freedom is 2, leading to the $12.5\sigma$
preference quoted in Section~\ref{sec:results}. This is the lowest significance
of all the DM templates compared to the RCG+NB template. The
preference for the RCG+NB template over the contracted NFW template, for example, is
$14.5\sigma$.

\subsection{Estimating the bulge-to-disk flux ratio} 
\label{app:BtoD}

\subsubsection{Initial estimate}

We estimate the bulge-to-disk ratio of GCE emission.  First, we estimate the
flux from detected MSPs within a local volume $D_\odot<3\mathrm{\,kpc}$.  To
identify pulsars as belonging to the local volume and to be millisecond pulsars
we compare all identified (PSR) and associated (psr) pulsars in the 3FGL
\citep{Acero:2015hja} to their counterparts in the ATNF catalog
\citep{Manchester:2004bp} and select those within 3 kpc from the Sun.
Furthermore, we impose a cut at 30 ms on the pulsation period.  In total we
identify 47 \gr\ pulsars with periods below 30 ms in the local volume. The
total flux from these sources is $1.2\times 10^{-9}\,
(6\times10^{-10})\mathrm{\,erg\,cm^{-2}\,s^{-1}}$ from $0.1-100\mathrm{\,GeV}$
(from $1-10\mathrm{\,GeV}$).

Next, we simulate a population of disk and bulge MSPs. For the spatial
distribution in the disk we use a Lorimer disk \citep{Lorimer:2003qc}, similar
to \cite{Fermi-LAT:2017yoi}.  The radial distribution is $\rho(r) \propto
r^{2.35}\exp{\left(-r/1.528\mathrm{\,kpc}\right)}$ and the height $\rho(z)
\propto \exp{\left(-|z|/z_0\right)}$ where we set the scale height $z_0 = 0.5$
kpc,~\cite{Levin:2013usa}.  Changing to $z_0 = 0.3\,(0.7)\mathrm{\,kpc}$
increases (decreases) the disk flux by 10\%.  For the bulge we assume a
spherically symmetric density profile that falls as $r^{-2.5}$ with a
hard-cutoff at $3 \mathrm{\,kpc}$~\citep{Calore:2014xka}.  Final results are
not expected to depend on the assumed bulge profile, since it is only used to
compare the expected flux from a bulge source to that of a disk source, and for
any bulge model the sources are located in the vicinity of the Galactic center.

We can obtain the disk flux from the local flux.  Using the simulation we can
estimate the number of sources in the local volume.
The fraction of local MSPs compared to the full population is
$f_\mathrm{local}=4.7\%$. In addition, we have to compare mean fluxes from
the local volume, disk and bulge.  For a given reference luminosity, $L_0$, the
mean flux can be calculated as 
\begin{equation}
  \left<S\right> \propto \int ds \frac{dN}{ds} \frac{L_0}{s^2},
\end{equation}
where $s$ is the distance along the line-of-sight and $\frac{dN}{ds}$ the
distribution of sources along the line-of-sight, which we obtain from the
simulation described earlier.  The disk flux can now be calculated
\begin{equation}
  S_\mathrm{disk} = \frac{S_\mathrm{local}}{f_\mathrm{local}}
  \frac{\left<S\right>_\mathrm{disk}}{\left<S\right>_\mathrm{local}}.
\end{equation}

Using the derived bulge flux from this paper we naively estimate $B/D$ flux
ratio to be $\sim 1.6$. Since the average bulge source is dimmer than the
average disk source, the luminosity ratio is even larger, $B/D\sim 4.1$. Note
that this estimate does not have a completeness correction for the disk.

\subsubsection{Adjusting the estimate for completeness}

In the above estimate we assumed that the flux from local MSPs is complete up
to a distance of 3 kpc. In the following we estimate the completeness and
modify our estimate accordingly.  \cite{Fermi-LAT:2017yoi} argue that down to
fluxes of $10^{-5}\mathrm{\,MeV\,cm^{-2}\,s^{-1}}$ in the $0.3-500$ GeV range
the efficiency for a pulsar in the inner Galaxy to end up being classified as a
pulsar candidate in the Second \Fermi~Inner Galaxy catalog (2FIG) is about 100\%.
This flux translates to $\sim2\times10^{-11}\mathrm{\,erg\,cm^{-2}\,s^{-1}}$
above 0.1 GeV.  Conservatively, we estimate the local number pulsars to be
complete down to $\sim4\times10^{-11}\mathrm{\,erg\,cm^{-2}\,s^{-1}}$. We find
a total of 8 sources with a flux equal or larger than this value.

To estimate the total flux from the local volume we simulate a population of
MSPs with a single power-law luminosity function:
$dN/dL\propto L^{-1.5}$ from $L_\mathrm{min}$ to 
$L_\mathrm{max}$, where $\left[L_\mathrm{min},
L_\mathrm{max}\right] = \left[10^{32}, 7\times 10^{34}\right]
\mathrm{\,erg\,s^{-2}}$ above $0.1\mathrm{\,GeV}$.  
The slope is motivated by 
$L\propto\dot E^\beta$ 
where $\dot E$ is the MSP spin-down power \citep{Strong:2006hf}.
The slope, $\beta$, 
is uncertain but appears close to unity \citep{Calore:2014oga, Venter:2014zea}.
Moreover, a bulge MSP population with this luminosity function has 
been shown to be 
able to explain the results from
\cite{Bartels:2015aea} for $L_\mathrm{max}=7\times 10^{34}
\mathrm{\,erg\,s^{-2}}$. 
Below we comment on the impact the assumed luminosity function
can have on the estimate of the disk flux.

Setting the number of
sources with $S\geq 4\times10^{-11}\mathrm{\,erg\,cm^{-2}\,s^{-1}}$ equal to
the observed number (8) we find a local flux of $\sim 2\times
10^{-9}\mathrm{\,erg\,cm^{-2}\,s^{-1}}$, almost a factor two larger than the
flux obtained from the 3FGL. The total flux is only mildly dependent on the
lower cutoff, $L_\mathrm{min}$.  Changing the slope of the luminosity function
to $\alpha=1.2\,(1.8)$ decreases (increases) the flux by about 10\%. However,
for the softer slope the flux becomes more sensitive to $L_\mathrm{min}$.

With these results we update our estimates for the bulge-to-disk ratio, taking
into account completeness, to a flux ratio of $B/D \sim 0.9$ and a luminosity
ratio of $B/D \sim 2.3$. This implies a $\sim10\times$ higher number of MSPs
per unit of stellar mass in the bulge compared to the disk \citep[for a disk
mass of $M_\mathrm{disk}=5.17 \times 10^{10}\mathrm{\,M_\odot}$] []{Licquia:2014rsa}.

\paragraph{Importance of the luminosity function} 
We highlight the importance of the assumed luminosity function in
the above estimate. Using the luminosity function of the form $dN/dL\propto
L^{-1.5}$ results in the local flux being estimated to be complete to within a
factor $\sim 2$. The reasoning behind this is that for this luminosity function
most of the flux comes from brightest sources, which are mostly resolved.
However, a different luminosity function, such as the one used in
\cite{Winter:2016wmy}, can yield a disk flux which is larger by an
order-of-magnitude.
In this case, the number of MSPs in the disk and bulge, when weighted by stellar mass, becomes comparable \citep{Eckner:ToAppear}.
The luminosity function assumed in \cite{Winter:2016wmy}
is given by a broken power-law,
\begin{equation}
  \frac{dN}{dL}\propto\left\{
	\begin{split}
      L^{-1.45},\,
      L<L_b\\
      L^{-2.86},\,
      L\geq L_b
    \end{split}
    \right.,
\end{equation}
with the break at $L_b = 8.7\times 10^{32} \mathrm{\,erg\,s^{-1}}$.  In this
case most of the flux is produced by sources with luminosities around the
break, $L_b$. The slope before the break is again consistent
with $L\propto \dot E$. However, the bright sources, which we observe,
are beyond the break and have a much steeper slope. A steeper slope
could for instance be explained through a faster spin-down rate
of the most luminous sources.
Since sources before the break would still be largely unresolved, 
this implies we
have only resolved the peak of the iceberg and that we should see many more sources
with slightly increased sensitivity. If the luminosity function from
\cite{Winter:2016wmy} represents the true luminosity function of MSPs the flux
in the resolved local sources would only contain $\mathcal{O}(10\%)$ of the
total local flux, meaning that the disk is expected to be much brighter. In
this case, the disk and bulge would roughly have an equal number of MSPs per
unit of stellar mass, an intriguing possibility which is studied
in detail in \citep{Eckner:ToAppear}.

\subsubsection{Comparing the emission from disk MSPs with ICS and gas emission}

To estimate the detectability of the MSP disk population 
component we compare its flux to the flux expected from ICS (comparable to a
thick disk) and $\pi^0$ emission (comparable to a thin disk).  The fluxes from
$1-10\mathrm{\,GeV}$ obtained for these components in the
$4.5^\circ\times180^\circ$ ROI (run \texttt{r5\_RCG\_NB}) are $7.8\times
10^{-8}\mathrm{\,erg\,cm^{-2}\,s^{-1}}$ ($\pi^0$) and $1.6\times
10^{-8}\mathrm{\,erg\,cm^{-2}\,s^{-1}}$.  We consider the disk flux (including
the completeness correction) within this ROI and compare to these
backgrounds.  The flux from MSPs in the disk in the 1-10 GeV band is $\sim8\%$ of
the ICS flux and $\sim 2\%$ of the $\pi^0$ flux.  Since the estimate for the
disk flux includes the resolved fraction, with sources in the local volume
already making up $\sim30\%$ of the total flux, the diffuse unresolved contribution
will be an even smaller fraction.  Consequently, the diffuse part of the
GeV emission from disk MSPs falls well within the uncertainties of this emission components and it
is unlikely to detect the disk as a separate component. 
However, we remind the reader that the estimate of the disk MSPs depends on the
assumed luminosity function, in case of a luminosity function similar to
that of \cite{Winter:2016wmy} the emission from disk MSPs becomes comparable to the ICS component.

In order to gauge the strength of the MSP disk \gr\ emission we performed a
fit including a Lorimer disk as described above with the spectrum fixed
to a stacked MSP spectrum. The normalization 
of this template remained close to zero (relative to 
other templates). Therefore, it appears
no strong MSP component in the disk is required.

\subsection{M31}
\label{sec:M31}

Recently, \cite{Ackermann:2017nya} detected \gr emission from M31 with a flux
of $(5.6\pm 0.6)\times 10^{-12}\mathrm{\,erg\,cm^{-2}\,s^{-1}}$ and found
marginal evidence for spatial extension.  Assuming M31 is similar to our
Milky Way we can estimate the predicted emission from M31 using the relation
between bulge stellar mass and the GeV emission.  Taking the distance to M31 to
be $785 \mathrm{\,kpc}$ and the stellar mass of the bulge $M_{\mathrm{M31}} =
(4.4-6.6)\times10^{10} \mathrm{\,M_\odot}$ \citep{Tamm:2012hw} we estimate that
the luminosity of M31 to be $\sim20\times$ that of the Milky Way bulge.  Thus,
assuming M31 is a Milky Way analog and that its \gr\ emission comes from the
bulge it would have $\sim 4\times$ more MSPs per unit of stellar mass. 
\cite{Eckner:ToAppear} arrive at the same conclusion, albeit through 
a different analysis in which they start from an estimation of the number of MSPs in the disk.

Assuming a similar ratios of GeV emission per unit stellar mass in the bulge
and disk of M31 as in those of the Milky Way, we estimate that $\sim90\%$ of
the potential GCE flux from M31 should come from the bulge and the remaining
$\sim 10\%$ from the disk.  Note that for the luminosity function where the
majority of the flux is produced by sources of intermediate luminosity, such as
that of \cite{Winter:2016wmy}, the disk of M31 can become equally bright, or
slightly brighter than the bulge, since in this case we expect the disk and
bulge to have a similar amount of MSPs per unit of stellar mass. 
\cite{Eckner:ToAppear} show that in this case MSP emission from the disk is 
only just below the detection threshold.
We note that
the radius of the bulge in M31 is only $\sim 0.1^\circ$ \citep{Tamm:2012hw}.
Consequently, the observed bulge emission of $0.4^\circ$ radius fit by a
uniform brightness disk from \cite{Ackermann:2017nya} appears large.
Future instruments with better spatial resolution
\citep[e.g.~e--ASTROGAM or
AMEGO\footnote{\url{https://asd.gsfc.nasa.gov/amego/index.html}}]
[]{DeAngelis:2016slk} could potentially better resolve the bulge
(also see \cite{Eckner:ToAppear}).


\section{Supplementary Information}

\label{sec:supp}
%

\begin{figure}[t]
  \centering
  \includegraphics[width=0.6\textwidth]{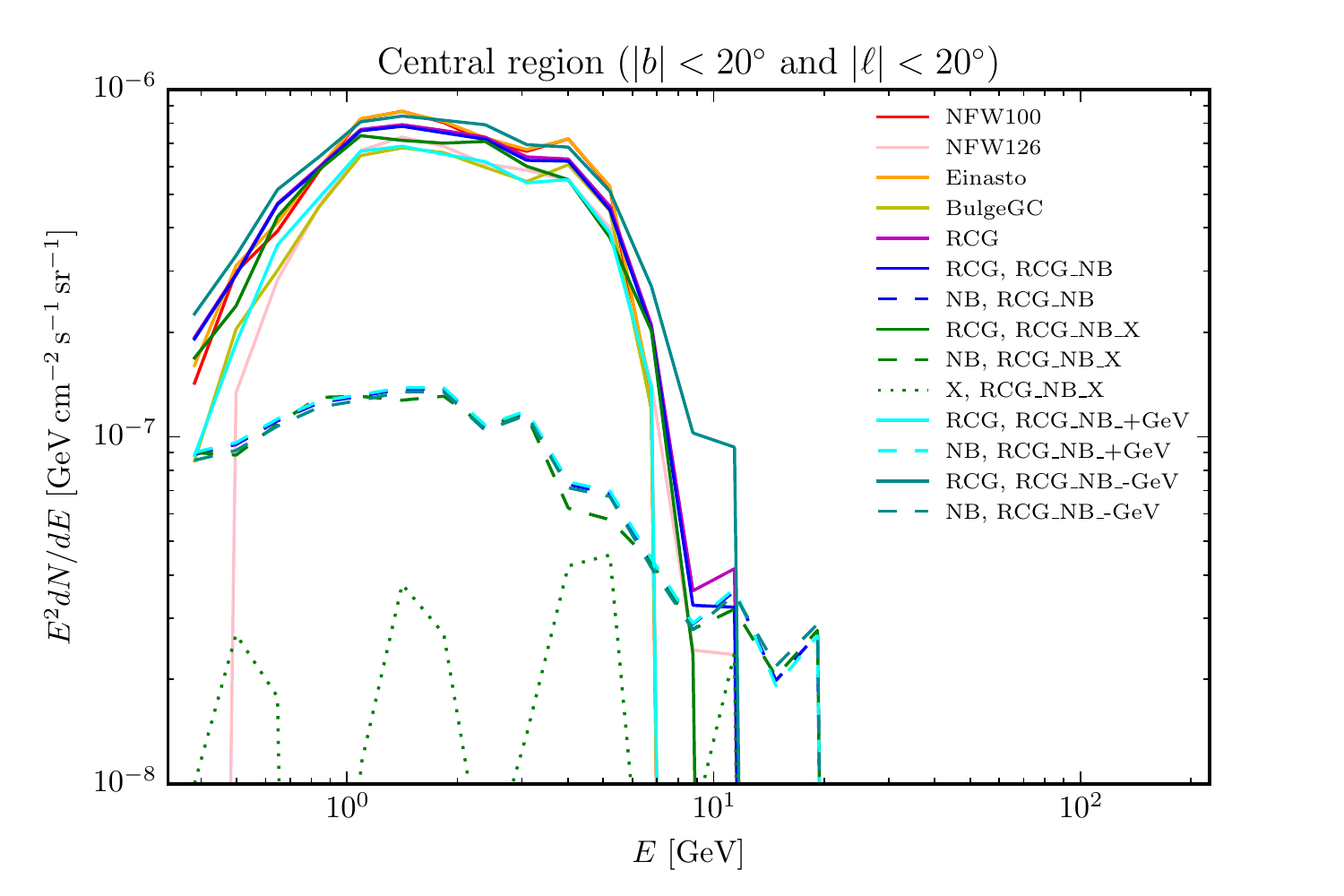}
  \caption{Recovered spectra for the GCE component of all runs in
    Tab.~\ref{tab:AH} starting with \texttt{r5}. If there are multiple GCE
    components in a single run (e.g., the RCG+NB template), the two components
    are plotted separately. Error bars not shown for clarity.}
  \label{fig:spectral_AllModels}
\end{figure}

\subsection{Overview of runs and \skyfact\ compared to traditional analyses}
\label{sec:template2skyfact}
\begin{table}[t]
  \centering
  \begin{tabular}{rl}
    \toprule
    Run ID & Comment \\
    \midrule
    \texttt{r5\_NFW126} &     Bulge spectrum from spatial DM template\\
    \texttt{r5\_NFW100} &     Bulge spectrum from spatial DM template\\
    \texttt{r5\_Einasto} &    Bulge spectrum from spatial DM template\\
    \texttt{r5\_BulgeGC} &    Bulge spectrum from spatial 511 keV template\\
    \texttt{r5\_RCG} &        Bulge spectrum from spatial stellar mass template\\
    \texttt{r5\_RCG\_NB} &    Bulge spectrum from spatial stellar mass template; two components\\
    \texttt{r5\_RCG\_NB\_X} & Bulge spectrum from spatial stellar mass template; three components\\
    \texttt{r5\_RCG\_NB\_+GeV} &    Softer Fermi bubble spectrum\\
    \texttt{r5\_RCG\_NB\_-GeV} &    Harder Fermi bubble spectrum\\
    \texttt{r5\_RCG\_NB\_pICS} &    Additional inner Galaxy ICS component with soft spectrum\\
    \texttt{r5\_RCG\_NB\_pICS2} &    Additional inner Galaxy ICS component with hard spectrum\\
    \midrule
    \texttt{A1} & Reproduction of \cite{Calore:2014xka}, with spatial DM template.\\
    \texttt{A2} & Reproduction of \cite{Calore:2014xka}, with spatial stellar mass template. \\
    \texttt{B1} & A1 gas/ICS templates changed to \run5 templates (1 gas template). \\
    \texttt{B2} & A2 gas/ICS templates changed to \run5 templates (1 gas template). \\
    \texttt{C1} & B1 with longitude extended to $|l|\leq90^\circ$. \\
    \texttt{C2} & B2 with longitude extended to $|l|\leq90^\circ$. \\
    \texttt{D1} & C1 with modulation as in \run5. \\
    \texttt{D2} & C2 with modulation as in \run5. \\
    \texttt{E1} & D1 with disk unmasked \\
    \texttt{E2} & D2 with disk unmasked \\
    \texttt{F1} & E1 with two gas rings used. \\
    \texttt{F2} & E2 with two gas rings used. \\
    \texttt{G1} & F1 with extended sources added in. \\
    \texttt{G2} & F2 with extended sources added in. \\
    \texttt{H1} & G1 with spectral bubble template. \\
    \texttt{H2} & G2 with spectral bubble template. \\
  \bottomrule
  \end{tabular}
  \caption{Overview of different runs that we used in the present paper.  All
    runs starting with \texttt{r5\_} are extensions or modifications of
    \texttt{run5} from \cite{Storm:2017arh}, where more details about the
    background modeling can be found.  \texttt{r5\_}  runs are also performed
    with the spectrum fixed to that obtained by \cite{McCann:2014dea} for
    stacked MSPs, in which case we append \texttt{\_msp} to the run ID.  Runs
    starting with \texttt{A}--\texttt{H} connect the analysis of
    \cite{Calore:2014xka} to \run5. The only difference between \run5 and
    \texttt{H} is that the gas template is broken up into 3 ring in the former and
    2 in the latter.  }
  \label{tab:runs}
\end{table}

Tab.~\ref{tab:runs} shows an overview of all \skyfact\ runs performed for this
analysis. Runs starting with \texttt{r5\_} are extensions or modifications of
\run5 from \cite{Storm:2017arh}. In Fig.~\ref{fig:spectral_AllModels}, we
show the best-fit spectra for the GCE components of all runs starting with
\texttt{r5\_} in Tab.~\ref{tab:runs}. For these runs, the recovered spectra
are quite similar, with less than 30\% variation in peak flux. The spectrum of
the X-shaped template is not well-recovered, since it is such a sub-dominant
component; the error bars are also large on this component. 

Runs starting with \texttt{A}--\texttt{H} in Tab.~\ref{tab:runs} connect the
analysis of \cite{Calore:2014xka} to \run5.  We used the latter to identify how
the current analysis differs from the previous results obtained with regular
template fitting.  To this end we changed the analysis assumptions step-by-step
until we well reproduce the results from \cite{Calore:2014xka}. 

In this section we elaborate on the runs
\texttt{A1}/\texttt{A2}--\texttt{H1}/\texttt{H2} as listed in Tab.~\ref{tab:runs}.  When referring to the runs we will now drop the appended
digit, and just mention that each runs was performed twice, ones for the NFW
(1) and ones for the bulge template (2).  The NFW template that is used has an
inner slope of $\gamma=1.26$ as in run $\texttt{r5\_NFW126}$ and the bulge
template includes the boxy bulge as traced by red clump giants and the nuclear
bulge, as in \texttt{r5\_RCG\_NB}. Note that the NB is fully masked when the
disk is masked.

\begin{table}[t]
    \centering
    \begin{tabular}{c|cc|c}
      \toprule
      & \multicolumn{2}{c|}{$-2\ln\mathcal{L}$} & $\Delta \chi^2$ \\
      {\bf Template}    & (1) NFW $\gamma=1.26$ & (2) RCG\_NB \\
      \midrule
      \texttt{A} & 147174.1 & 147486.0 & -311.9\\
      \texttt{B} & 165359.2 & 167419.6 & -2060.4\\
      \texttt{C} & 718013.4 & 721344.0 & -3330.6\\
      \texttt{D} & 562568.4 & 562995.2 & -426.8\\
      \texttt{E} & 655669.2 & 654782.6 & 886.6\\
      \texttt{F} & 655113.2 & 654947.4 & 165.8\\
      \texttt{G} & 651279.8 & 651022.9 & 256.9\\
      \texttt{H} & 648635.9 & 648484.5 & 151.4\\
      \bottomrule
      \end{tabular}
    \caption{$-2\ln\mathcal{L}$ values for the various runs connecting
      \cite{Calore:2014xka} to \run5.  $\Delta \chi^2 =
      -2(\ln\mathcal{L}_\mathrm{NFW} - \ln\mathcal{L}_\mathrm{RCG\_NB})$, where
      RCG\_NB refers to the combination of boxy bulge and nuclear bulge.  }
    \label{tab:AH}
\end{table}
Run \texttt{A} mostly reproduces \texttt{Model F} from \cite{Calore:2014xka}.
It uses a smaller, $40^\circ\times40^\circ$ ROI, the \texttt{GALPROP} gas and
ICS templates from \texttt{Model F}, a uniform spatial bubble template
\citep{Su:2010qj} and the Galactic disk is masked $|b|\geq 2^\circ$. The main
difference is the point source treatment: runs \texttt{A--H} include all the
3FGL sources in the fit with free normalizations (same constraints as
\texttt{run5} in \citealt{Storm:2017arh}), whereas \citet{Calore:2014xka} mask
point sources.  For this run, the NFW template is preferred over the bulge
template. We show the difference in fit quality, $\Delta \chi^2$, between the
the NFW and bulge runs in Tab.~\ref{tab:AH}.  Step-by-step we release the
constraints until we reproduce \run5.  In model \texttt{B} we change the gas
and ICS templates to the ones used in \run5; however, rather than using 3 rings
for the $\pi_0$ template only a single template is used.  Changing the gas
template does not affect the preference for the NFW template.  Run \texttt{C}
extends the ROI to include the full latitude range used in the rest of this
work, $|l|\leq 90^\circ$. The NFW profile still is preferred.

The largest change in the $\Delta \chi^2$ occurs between run \texttt{C} and
\texttt{D}, where modulation parameters for the spatial templates are switched
on. The NFW template is still formally preferred, but the difference is
drastically reduced. If only the Poisson term of the likelihood is considered,
the switch in preference for NFW to RCG+NB actually occurs here between runs
\texttt{C} and \texttt{D}, but the regularization term in run \texttt{D} is
large enough so that for the total likelihood, the NFW template is still
preferred.  Unmasking the disk in run \texttt{E} results in an overall
preference for the RCG+NB template over the NFW. Next, the $\pi_0$ template is
broken up into two rings in run \texttt{F}.  Extended sources are added in run
\texttt{G}. There is still a preference for the bulge template.  Finally, we
change the template for the bubbles from a uniform spatial template to a
spectral template and thus recover \run5 in \texttt{H}. Run \texttt{H1}
(\texttt{H2}) is identical to \texttt{r5\_NFW} (\texttt{r5\_RCG\_NB}) except
that the latter uses 3 gas rings. The choice of 2 or 3 gas rings does not
affect the fit to the GCE.

\subsection{Relevance of modulation parameter ranges}
\label{sec:mod}

A critical component of the analysis, and in fact part of the reason why we
obtain qualitatively different results than previous studies, is the inclusion
of nuisance parameters in the various model templates.  We here briefly
summarize the main assumptions on the nuisance parameters that we made for our
reference model.  Full details can be found in \citet{Storm:2017arh}.
Variations at the levels smaller than or equal to the constraints described
above would be not surprising, and are of the size of actual residuals that one
can find in standard template analyses.

Robust results can only be obtained if they do not critically depend on the
choices made.  To test this, we increase or decrease the allowed range of the
spatial or spectral nuisance parameters by a factor four, for each of the
diffuse (the ICS component plus the three gas rings) components separately, as
well as for the 3FGL sources.  We find that the preference for the RCG+NB model
over the Einasto model is rather stable under changes.  More specifically, the
$\Delta \chi^2$ changes to 364 (72) when the spatial nuisance
parameter range of the outer gas ring is increased (decreased) by a factor of
4. In all other cases, the variation is smaller.  We checked that the inferred
best-fit spectra for the RCG and NB components does not vary by more than
$30\%$ or a factor of two, respectively.    We also checked that increasing the
spectral freedom of the three gas rings or the ICS component individually has
no noticeable effect on our results.

\subsection{Relevance of the size of the ROI}
\label{sec:ROI}

We find that the large ROI that we use in our analysis ($|b|<20^\circ$ and
$|\ell|<90^\circ$) plays an important role for obtaining robust and stable
results.  The central point is that the best discriminator for the various
diffuse model components is their spatial extent along the Galactic plane and
towards high latitudes.  This discriminator is lost if the analysis is limited to a too-small ROI.  Instead, the results start then to depend on specific
small-scale shape differences between the components, which are usually hard to
model and susceptible to a large range of systematic uncertainties. This becomes particularly problematic if the signal of interest has the same spatial extent as the ROI.

For instance, the different gas components overlap in the inner Galaxy along
the Galactic disk (say, $|b|<5^\circ$, $|\ell|<20^\circ$), and are hard to
discriminate in a fit that is limited to an ROI of comparable size.  However,
the various components differ -- by construction -- in their longitudinal
extent.  Including a larger portion of the Galactic disk, as we do in our
analysis, provides enough leverage to disentangle the components and constrain
their energy spectrum and normalization.  At the same time, the modulation
parameters that we introduced in our analysis allow to account for inaccuracies
of the modeled components at smaller scales.  A similar argument holds for our
treatment of the Fermi bubbles (see
Sec.~\ref{sec:bubbles}) and ICS emission.  

\medskip

\paragraph{The X-shaped bulge.}
The above effects explain why we can not confirm the recent claims by
\cite{Macias:2016nev} that the GCE is best described by an
X-shaped bulge template.  In fact, we find that we can reproduce the results
from this paper if we (a) restrict our analysis to the same ROI (a
$15^\circ\times15^\circ$ region around the GC), (b) allow complete spectral
freedom and (c) fix the spatial templates.  Under these conditions, we find
indeed that the X-shaped bulge is preferred over the RCG+NB template, with a
similar best-fit spectrum as shown in \cite{Macias:2016nev}.  
However, from our analysis it appears that most of the emission previously attributed to the GCE is \textit{falsely}
absorbed by the gas templates when using the smaller ROI.
We find that such a fit leads to
results inconsistent with the data as soon as one increases the size of the
ROI.

One might argue that a small ROI should lead to more robust results, by
increasing statistical error bars and making the results less dependent on a
mismodeling of diffuse Galactic emission at large scales.  However, this is only strictly true if the shape of all model components is accurately
known.  Mismodeling will introduce biases, which play an increasingly large
role if the most prominent differences between the signal of interest and the
background components are removed by using a small ROI.

A potential caveat when comparing to the results of \cite{Macias:2016nev} is that they apply newly developed 
hydrodynamical gas maps which
they find to be highly preferred over the interpolated gas maps
we use in our analysis. However, since we are able to reproduce their
main results also when using interpolated gas maps, 
namely a preference for an X-shaped bulge with a similar 
spectrum as found in \cite{Macias:2016nev}, we do not expect the
hydrodynamical gas maps to be driving the preference for an X-shaped
component. Rather, we find that when connecting our analysis to theirs the smaller ROI drives the difference, which as argued above 
leads to inconsistencies when extrapolating the templates outside of
their fit region.

\subsection{Fermi bubbles and the role of star formation in the CMZ}
\label{sec:bubbles}

Using H.E.S.S. and \Fermi-LAT data, \cite{2017MNRAS.467.4622J} estimate the
supernova recurrence time in the CMZ to be $\sim2500\rm\,yrs$.  Assuming 10\% of
the kinetic energy goes into cosmic rays, and 1\% into electrons, this corresponds to
an average injection of $\sim 10^{37}\mathrm{\,erg\,s^{-1}}$ into cosmic-ray electrons.
This is comparable to the luminosity of the GCE component in our fits
($\sim1.7\times 10^{37}\mathrm{\,erg\,s^{-1}}$).  However, it is generally
expected that a fraction of the electron cosmic-ray energy is lost at radio frequencies
by synchrotron radiation.  Furthermore, the expected \gr\ spectrum would
be most likely hard, comparable to the \Fermi\ Bubbles spectrum or the spectrum of
our ICS template.  In our analysis, we expect that this component
manifests as low-latitude component of the \Fermi\ Bubbles.  Interestingly, we
find that total luminosity of our low-latitude \Fermi\ Bubbles component is $\sim
3.6\times 10^{36}\mathrm{\,erg\,s^{-1}}$
(for a $10^\circ\times 10^\circ$ ROI around the GC integrated over all energies, run \texttt{r5\_RCG\_NB}). A possible connection to star formation in the CMZ will be explored elsewhere.

The spectrum of the GCE differs significantly from the high-latitude \Fermi\
Bubbles spectrum \citep[e.g.~][]{Huang:2015rlu}.  However, the emission that
should be associated with the \Fermi\ Bubbles at low energies is unknown.  We
test the robustness of our results against different \Fermi\ Bubbles
characterizations.  For this purpose we reanalyze \texttt{r5\_RCG\_NB} with a
hardened/softened bubble spectrum, by multiplying its original spectrum by a
power law of $E^{\pm0.05}$ (\texttt{r5\_RCG\_NB\_$\pm$GeV}).  As seen in
Fig.~\ref{fig:spectral_AllModels}, the spectra of the RCG and NB components
are very similar to those in the run with the original bubble spectrum,
changing by less than $20\%$.  In addition, as discussed in
Section~\ref{sec:template2skyfact}, we implemented a fixed uniform morphology
for the \Fermi\ Bubbles as in \cite{Calore:2014xka}, in which case the GCE component is
increased by $\sim20\%$.  Lastly, test models where we allow an additional
spectral component in the inner $10^\circ\times 10^\circ$ region of the
Galactic center, with completely free morphology, and a spectrum either fixed
to the ICS component of our main runs (which is significantly softer than the
\Fermi\ Bubbles spectrum), or to the spectrum of the \Fermi\ Bubbles multiplied
by $E^{0.1}$, which is significantly harder.  In the first case
(\texttt{r5\_GCE\_NB\_pICS}), this component does not pick up a significant
amount of the inner Galaxy emission, but in the second case
(\texttt{r5\_GCE\_NB\_pICS2}) it absorbs much of the enhanced bubble emission
that we see in our main runs.  In both cases, the GCE flux is not affected much
(the RCG component by less than 20\%, the NB is barely affected).  This
suggests that the freedom introduced with of our \Fermi\ Bubbles template is
already enough to account, together with the GCE, for most of the inner Galaxy
emission.

\subsection{The role of point sources}

A recent dedicated search for point sources in the inner Galaxy was performed
by the Fermi Collaboration and resulted in the publication of the 2FIG catalog
\citep{Fermi-LAT:2017yoi}. We test the effect of including these additional
sources on the GCE in our runs by including all 2FIG sources that are not in
the 3FGL catalog and not found in clusters near other known point or extended
sources. The spectra and positions are taken from the 2FIG catalog and the same
modulation parameters as for the 3FGL sources are used. We find that the
adding the 2FIG sources does not change the preference for the RCG+NB
templates over the DM templates. The RCG+NB template is still preferred over
the Einasto profile by a $\Delta \chi^2>100$. The GCE fluxes are reduced by $<5\%$ to 15\% at most when including the 2FIG
sources. The one exception is the X-shaped template; the flux of this template
is reduced by 50\%. However, this component is already so sub-dominant it has
little effect on the overall quality of the fit. 

\end{document}